# FLOW AND THERMAL FIELDS INVESTIGATION IN DIVERGENT MICRO/NANO CHANNELS

**Amin Ebrahimi**
High-Performance Computing (HPC) Laboratory,
Dept. of Mechanical Eng., Faculty of Eng.,
Ferdowsi University of Mashhad, Iran.

***Ehsan Roohi**
High-Performance Computing (HPC) Laboratory,
Dept. of Mechanical Eng., Faculty of Eng.,
Ferdowsi University of Mashhad, Iran.

## ABSTRACT
The present work is related to the study of the nitrogen gas flow through diverging micro/nano-channels. The direct simulation Monte-Carlo (DSMC) method has been used to study the flow. The Simplified Bernoulli Trials (SBT) collision scheme has been employed to reduce the computational costs and required amounts of the computer resources. The effects of various divergence angles on flow and thermal fields have been studied for different Knudsen numbers in late-slip and early-transition regimes. The inlet-to-outlet pressure ratio has been set to 2.5 for micro/nano-channels with a uniform constant wall temperature. By analyzing the numerical results no flow separation has been found due to slip at the wall which is different than flow behavior in continuum regime. The results indicate that the viscous component has a relatively large contribution in the overall pressure drop and flow behavior. It observed that for low divergence angles the effects of pressure forces dominate the effects of shear stress and divergence angle and cause the flow to accelerate along the channel while by increasing the divergence angle and therefore the effects of flow expansion, the flow decelerates along the channel. The mass flow rate through channel increases by increasing the divergence angle. Cold-to-hot heat transfer has been observed in diverging micro/nano-channels. In order to investigate the thermal behavior in diverging micro/nano-channels, the results have been compared to weakly non-linear constitutive laws derived from Boltzmann's equation.

## INTRODUCTION
Improving the performance and decreasing the size of equipment in micro-electronics are the most important incitation factors for sciences and economic developments in the last fifty years. Nowadays, microsystems has application in different industries such as micro-pumps, micro-turbines, micro-valves, micro-accelerometer, micro-actuators, micro-heat exchangers and micro propulsion systems for micro/nano-satellites. Channels at micro/nano scales are one of the most essential parts in these equipments. Many researchers have focused on flow hydro/thermal field behavior in micro/nano-channels [1-4].

Expansion or contraction phenomenon can be found in many micro/nano-systems. Rathakrishnan and Sreekanth [5] studied rarefied gas flow (Kn = 0.0026 – 1.75, Nitrogen) through circular tube with sudden increase in cross sectional area. They noted that in the transition regime, the pressure ratio and length to diameter ratio of the passage strongly influence the discharge through sudden enlargements. In a recent experimental study involving flow through a sudden expansion, Varade et al. [6] observed a discontinuity in the slope of pressure and absence of flow separation at the junction; in the slip regime. These measurements are qualitatively similar to the two-dimensional planar simulations of Agrawal et al. [7]. Lee et al. [8] in an experimental study on gas flow through microchannels connected through diverging section observed that the mass flow rate decreases and the pressure loss increases with increasing included angle of the transition section. Varade et al. [9] focused on investigating low mach number slip flow behavior through diverging microchannel at high inlet-to-outlet pressure ratios. They found three different flow behaviors in diverging micro-channels. Varade et al. [10] studied the gaseous flow in a converging microchannel in slip regime.

The Direct Simulation Monte-Carlo (DSMC) method is a particle-based algorithm for simulating rarefied gas flows [11] and is valid for the investigation of gas flows in regimes ranging from continuum to free-molecular conditions. The degree of gas rarefaction is determined by the Knudsen number ($Kn=\lambda/D$), which is defined as the ratio of mean free path ($\lambda$) of gas molecules to a characteristic length, D. The rarefaction regimes can be generally categorized as slip ($0.001<Kn<0.1$), transition ($0.1<Kn<10$), and free molecular ($Kn>10$) ones. Recently, Stefanov [12] proposed the simplified Bernoulli trials (SBT) collision scheme as an alternative to NTC scheme that avoids the repeated collision in cells and permits simulations using a much smaller mean number of particles per cell (PPC). It is shown that SBT scheme could obtain an accurate solution even with PPC=2 and even less while the NTC scheme requires a PPC around 15~20 for typical test cases [13]. Detailed information about DSMC method and SBT collision scheme can be found in [13, 14]. The DSMC code used in the present work has been implemented in OpenFOAM. This solver, has been validated for a variety of benchmark cases [15]. In our extended DSMC solver the collisions between particles are simulated using Variable Hard Sphere (VHS) collision model and Larsen–Borgnakke internal energy redistribution model [11].

It can be noted that the rarefied flow and thermal field characteristics through diverging micro/nano-channels were not considered in previous studies appropriately. More specifically, the possibility of the cold to hot heat transfer in the diverging micro/nano-channels and reasons behind it was not reported in the literature. The objectives of this work are to investigate the gas flow behavior through diverging micro/nano-channels with different configurations at different rarefaction regimes and to highlight significant differences with respect to continuum flow behavior.

**PROBLEM DESCRIPTION**
The Poiseuille flow of nitrogen gas through a divergent nano-channel is investigated in this paper. The schematic diagram of the geometry is shown in figure 1, where the inlet height of channel ($H_{in}$) is 400nm, and the length (L) is 20H. The divergence angle ($H_{out}/H_{in}$) is considered as a parameter to study and is equal to 1.0, 1.5, 3.5 and 7.0. The wall and inlet gas temperatures are all equal at 300 K. The inlet to outlet pressure ratio (PR) is 2.5. The Knudsen number based on inlet height ($H_{in}$) is set to 0.05, 0.1, 0.25 and 0.50. Only one half of the channel is used for simulation due to the symmetry of the channel. The diffuse reflection model is selected to treat the solid walls. The solution is continued even when the values of mass flow rate at the inlet and outlet become equal decrease the statistical fluctuations in flow parameters.

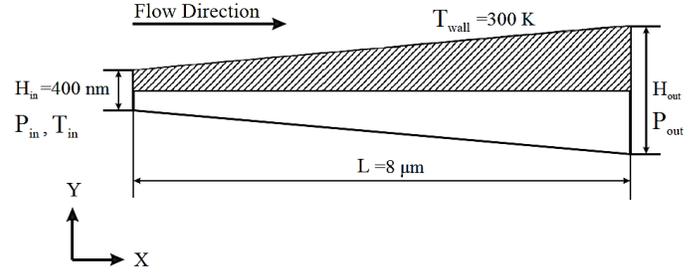

**Fig 1**: Schematic diagram of divergent micro/nano-channel

**RESULTS AND DISCUSSIONS**
To finalize the suitable grid sizes and number of DSMC simulator particles in each cell, grid and particle independence tests are performed. A grid with cell sizes of $\lambda$ and $\lambda/3$ in stream-wise and normal direction, respectively, is selected after grid study. Additionally, the simulations are initialized with PPC=5.

To validate our extended DSMC solver a micro-channel Poiseuille flow of nitrogen gas is used. This benchmark includes a channel of height 0.4 μm, and 2 μm lengths, meshed with 100×60 computational cells. The surface and inlet gas temperatures are set to 300K. The outlet pressure is 101325 Pa, and the inlet to outlet pressure ratio is 2.5. The Knudsen number at the inlet is 0.055 and 0.123 at the exit. These simulations contained two and five DSMC simulator particles in each cell and was solved in parallel on two processors. The results are compared with previous numerical work of the same case from Liou and Fang [16], Roohi et al. [3] and White et al. [1] as well as first and second order analytical slip solution and presented in figure 2 and a good agreement is achieved.

After ensuring the proper functioning the developed solver, the rarefied flow and thermal field characteristics through diverging micro/nano-channels in the late-slip and early-transition Knudsen number regimes are studied here. Figure 3 shows the normalized pressure distributions for different channel configurations when $Kn_{in}=0.1$. It is observed that an increase in divergence angle would result in a decrease in the slope of pressure distribution. It should be noted that compressibility increases the pressure nonlinearity while the rarefaction causes more linearity for the pressure distribution. For the divergent channels the expansion effects, due to the increase of the cross-section flow area, become more dominant than rarefaction effects. Figure 4 represents the local Mach number of different channel configurations when $Kn_{in}=0.1$. It observed that for low divergence angles the effects of pressure forces dominate the effects of shear stress and divergence angle and cause the flow to accelerate along the channel while by increasing the divergence angle and therefore the effects of flow expansion, the flow decelerates along the channel. It is in consistence with pressure distribution once by decreasing the pressure along the channel the velocity in the channel will increase. It is also seen that by increasing the divergence angle and increasing the flow Mach number the kinetic energy of the flow is increased which results in lower internal energy and fluid temperature. The variations of temperature are illustrated in figure 5 for different divergence angles at $Kn_{in}=0.1$. At a region near the channel outlet the temperature is dropped suddenly which may be attributed to effects of expansion cooling.

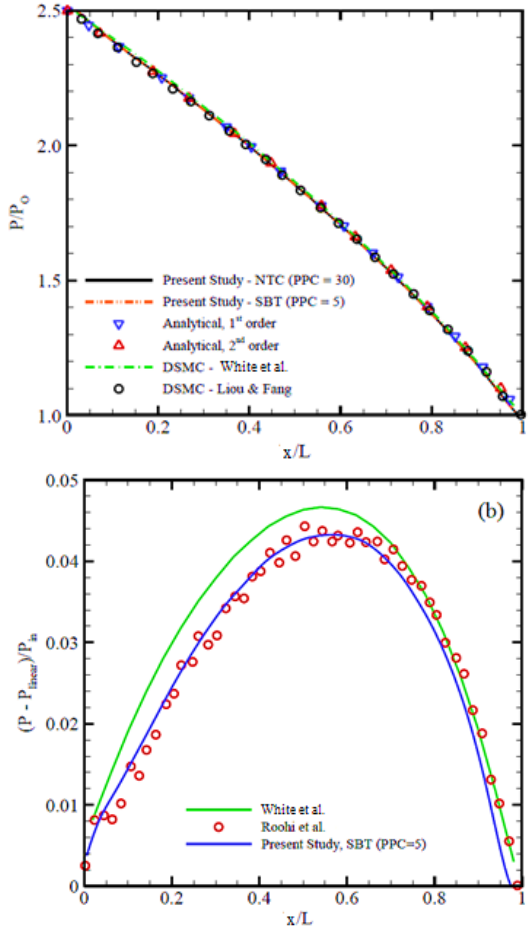

**Fig 2**: Comparisons with available numerical and analytical works for the micro-Poiseuille flow geometry for (a) stream-wise centerline pressure distribution, and (b) deviation of this distribution from a linear profile.

Figure 6 shows the normalized pressure distributions of the channel with $H_{out}/H_{in}=3.5$ for different Knudsen numbers. As expected, non-linear pressure profiles have been found, and the degree of non-linearity decreases with increasing the Knudsen number as rarefaction effects begin to dominate the expansibility effects. Figure 7 indicates the local Mach number of the channel with $H_{out}/H_{in}=3.5$ for different Knudsen numbers. It is clearly observable that as the Knudsen number increases, the centerline Mach number along the divergent nano-channel decreases. Figure 8 shows the centerline temperature along the diverging microchannel for different Knudsen numbers. Increasing the Knudsen number results in approaching an isotherm process along the diverging micro/nano-channels. A simple zero order analysis is carried out by Gavasane et al. [17] to calculate the outlet temperature by solving the conservation of mass, momentum and energy.

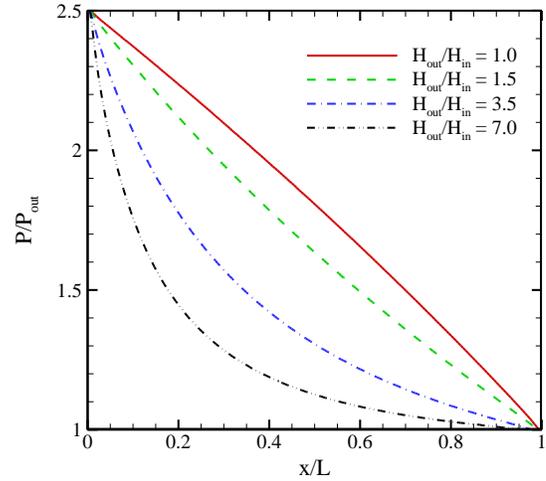

**Fig 3**: Stream-wise center-line pressure profiles. ($Kn_{in}=0.1$)

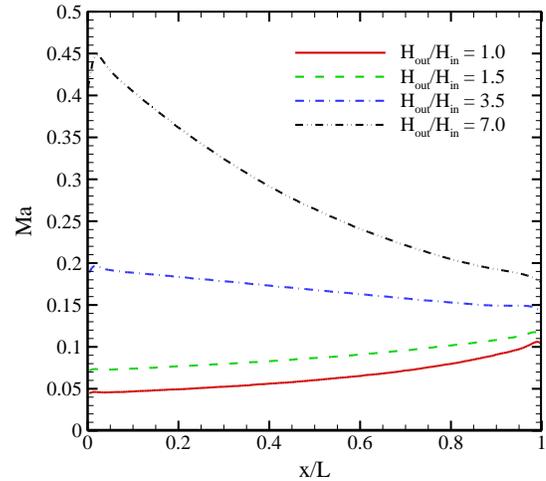

**Fig 4**: Stream-wise center-line Mach number profiles. ($Kn_{in}=0.1$)

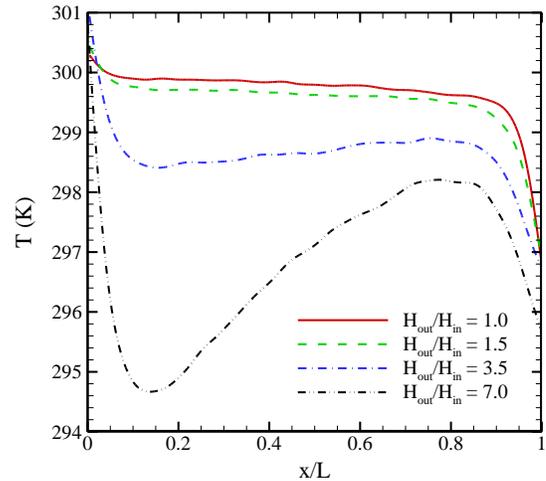

**Fig 5**: Stream-wise center-line temperature profiles. ($Kn_{in}=0.1$)

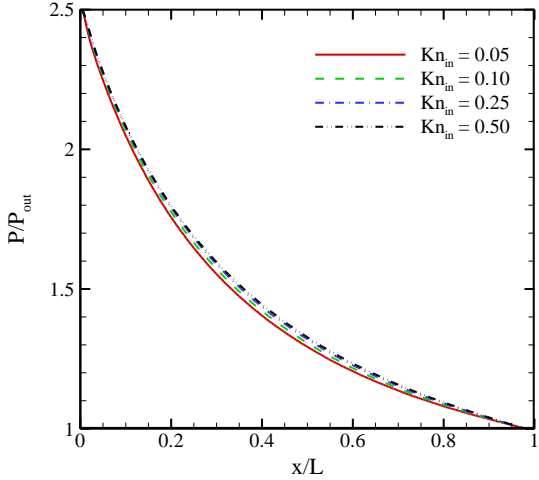

**Fig 6**: Stream-wise center-line pressure profiles. ($H_{out}/H_{in}$=3.5)

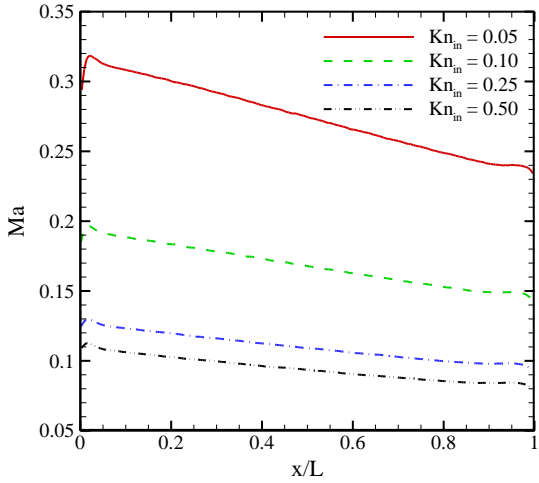

**Fig 7**: Stream-wise center-line Mach number profiles. ($H_{out}/H_{in}$=3.5)

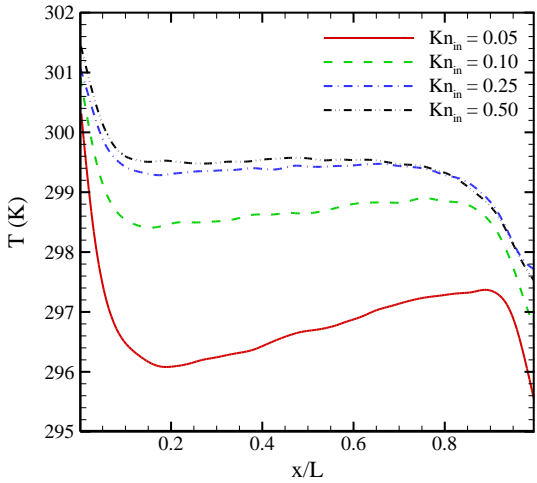

**Fig 8**: Stream-wise center-line temperature profiles. ($H_{out}/H_{in}$=3.5)

To investigate the mass flow rate through micro/nano-diverging channels the mass fluxes are plotted in figure 9. Each of the data points has been normalized with the mass flux from the corresponding straight micro/nano-channel. Figure 9 shows a flow enhancement by increasing the divergence angle of the channel in which this enhancement is more detectable at lower Knudsen numbers and higher divergence angles. For instance, at $Kn_{out}$=0.05 the mass flux is up to 11 fold higher in channel with $H_{out}/H_{in}$=7.0 compared to straight channel and is nonlinearly related to the $1/Kn_{out}$. In divergent micro-channels the shear stress is higher than that of a straight micro-channel for a same Knudsen number.

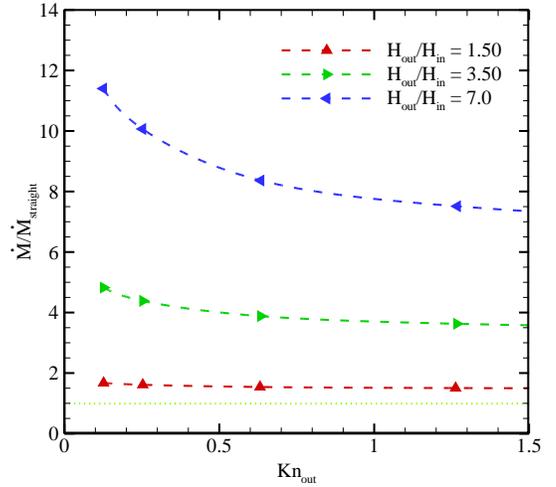

**Fig 9**: Mass flux measurements in diverging micro/nano-channels.

Figure 10 shows that cold to hot heat transfer occurs higher Knudsen numbers like $Kn_{in}$=0. 5, i.e., heat lines move back from the low temperature region toward the hot region. As can be seen in figure 10, temperature decreases suddenly at x/L=20 because of flow expansion and a cold region appears. Heat lines originating from micro-channel wall enters this area and but they are deflected toward the hotter region near the inlet. To investigate this behavior, more advanced constitutive laws should be considered, as Fourier's law is unable to estimate the behavior of the gas at the rarefied flow condition. Here, the weakly nonlinear form of the Boltzmann equation derived by Sone [18] has been used to understand the gas behavior. These set of equations take into account the dependency of the gas thermal conductivity on the temperature. As the Knudsen number and temperature variations are small in the cases studied here, the use of the constitutive laws of weakly nonlinear equations derived by Sone is justified. In these equations, the heat flow vector $Q_i$ is given as follows:

$$Q_{iS} = Kn^{**}Q_{iS1} + Kn^{**2}Q_{iS2} + Kn^{**3}Q_{iS3} \qquad (1)$$

$$Q_{iS1} = 0 \qquad (2)$$

$$Q_{iS2} = -\frac{5}{4}\gamma_2 \frac{\partial \tau_s^{**}}{\partial x_i} \qquad (3)$$

$$Q_{iS3} = -\frac{5}{4}\gamma_2 \frac{\partial \tau_s^{**}}{\partial x_i} - \frac{5}{4}\gamma_5 \tau_s^{**}\frac{\partial \tau_s^{**}}{\partial x_i} + \frac{1}{2}\gamma_3 \frac{\partial^2 u_{iS1}^{**}}{\partial x_j^2} \quad (4)$$

The employed parameters are defined as follow:

$$Kn^{**} = Kn\frac{\sqrt{\pi}}{2} \quad (5)$$

$$\tau^{**} = \frac{T - T_0}{T_0}, u = \frac{u^{**}}{\sqrt{2RT}}, x = \frac{x}{L} \quad (6)$$

The coefficients used in Eq. (7) are set as $\gamma_2$=1.9222, $\gamma_3$=1.9479 and $\gamma_5$=1.9611 [18]. The interplay between Fourier term $(-\partial \tau_s^{**}/\partial x_i)$ in $Q_{iS2}$ and $\partial^2 u_{iS1}^{**}/\partial x_j^2$ in $Q_{iS3}$ determines the heat paths behavior because the magnitude of $\partial^2 u_{iS1}^{**}/\partial x_j^2$ term is more dominant than others in $Q_{iS3}$. It is found that the second derivative of the velocity is directed toward the inlet. Figure 10(b) shows the overall heat lines (Eq. 1), which matches with the DSMC results. It can be concluded that in cases where there is no difference between the inlet temperature and wall temperature, the temperature gradient is small and, therefore, the effect of Fourier term becomes limited. The deviations of Sone formulation from DSMC results will increase with increasing the Knudsen number; therefore, more complicated formulation is required to obtain more accurate results.

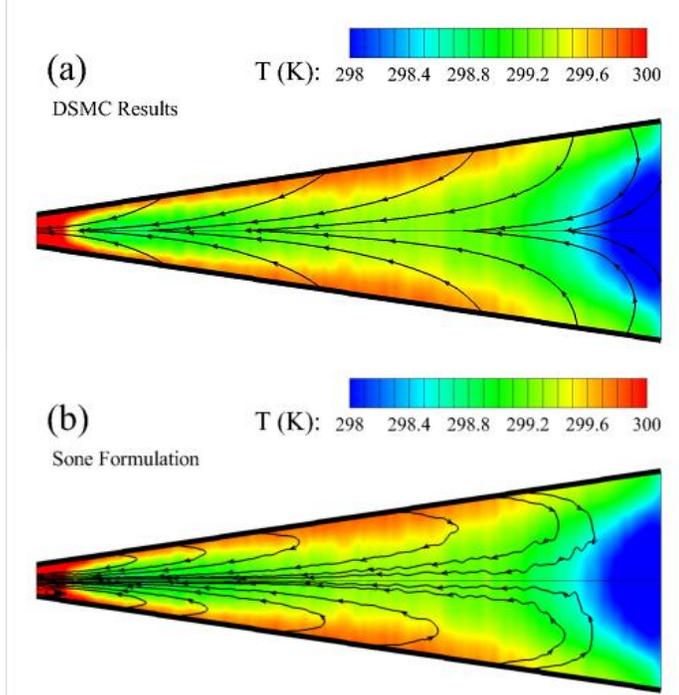

**Fig 10**: Temperature contours with heat lines, $H_{out}/H_{in}$=7.0, $Kn_{in}$=0.5. (a) DSMC results, (b) Sone formulation

## CONCLUSION
The DSMC method with the SBT collision scheme used to simulate subsonic Poiseuille flow of nitrogen gas through divergent micro/nano-channels. These simulations are implemented in the framework of an open-source flow solver, OpenFOAM, and was solved in parallel on four processors. The results presented for different Knudsen numbers and divergence angles at a pressure ratio of 2.5. It is observed that by increasing the divergence angle and Knudsen number the compressibility effect is weakened. Furthermore, it is found that due to slip at the wall of the divergent micro/nano-channels there is no reversal flow. The mass flow rate of the divergent micro/nano-channels compared with straight micro/nano-channels and it is found that the mass flow rate increases by increasing the divergence angle and decreasing the Knudsen number. Channels with higher divergence angles results higher mass flow rates at a fixed pressure drop. Cold to hot heat transfer has been observed. By decreasing the Knudsen number, the cold to hot heat transfer is weakened due to stronger Fourier term.

## NOMENCLATURE
H    Height
Kn   Knudsen number
L    Length
Ma   Mach number
P    Pressure
T    Temperature